\documentclass[trackchanges,twocolumn]{aastex701}

\begin{document}

\title{The Spatially-Resolved Recent Star Formation History of the M81 Disk}

\author[0000-0002-8406-0136]{Vaishnav V. Rao}
\affiliation{Department of Astronomy, University of Michigan, 1085 S. University Ave, Ann Arbor, MI 48109-1107, USA}
\email[show]{vvrao@umich.edu}

\author[0000-0002-5564-9873]{Eric F.\ Bell}
\affiliation{Department of Astronomy, University of Michigan, 1085 S. University Ave, Ann Arbor, MI 48109-1107, USA}
\email{ericbell@umich.edu}

\author[0000-0003-2599-7524]{Adam Smercina}\thanks{NHFP Hubble Fellow}
\affiliation{Space Telescope Science Institute, 3700 San Martin Dr., Baltimore, MD 21218, USA}
\email{asmercina@stsci.edu}

\author[0000-0001-6320-2230]{Tobin M. Wainer}
\affiliation{Department of Astronomy, University of Washington, Box 351580, Seattle, WA 98195, USA}
\email{tobinw@uw.edu}

\author[0000-0002-8084-8612]{Andrew Dolphin}
\affiliation{Raytheon, Tucson, AZ 85756, USA}
\affiliation{Steward Observatory, University of Arizona, Tucson, AZ 85721, USA}
\email{adolphin@rtx.com}

\author[0000-0002-7502-0597]{Benjamin Williams}
\affiliation{Department of Astronomy, University of Washington, Box 351580, Seattle, WA 98195-1580, USA}
\email{benw1@uw.edu}

\author[0000-0002-1264-2006]{Julianne J.\ Dalcanton}
\affiliation{Center for Computational Astrophysics, Flatiron Institute, 162 Fifth Avenue, New York, NY 10010, USA}
\affiliation{Department of Astronomy, University of Washington, Box 351580, Seattle, WA 98195-1580, USA}
\email{jdalcanton@flatironinstitute.org}

\begin{abstract}
The nearby M81 galaxy---a tidally interacting $L^*$ spiral disk---offers a unique perspective on the interplay of secular and external processes in shaping galaxy properties. We present the most detailed, panoramic map of M81's stellar populations to date, using archival Hubble Space Telescope ACS photometry from the ANGST survey spanning the full optical extent of its disk. From color-magnitude diagram fitting, we recover spatially-resolved star formation histories (SFHs) over the last $\sim180\,$Myr. M81's global SFH is remarkably steady in the last $\sim180\,$Myr with slight enhancements seen $\lesssim15\,$Myr ago. The average SFR surface density rises toward the center before falling off within the innermost $\sim4\,$kpc. The disk is also notably lopsided, with the eastern half having formed $\sim50\%$ more stars than the western half over the same area---a trend missed by FUV-based SFR indicators due to dust systematics. Finally, the long-term SFHs of both M81 and M82 point to a pericenter passage $\sim1\,$Gyr ago. By identifying M81-M82 analogs in the TNG50 cosmological simulation, we rule out the more recent pericenter times of $150-220\,$Myr predicted by some works, as such passages are inconsistent with M81's undisturbed SFH over the last $\sim180\,$Myr. With this work, M81 becomes only the third $L^*$ spiral---after M31 and NGC 6946---to have its SFH spatially resolved across large areas of its disk.

\end{abstract}



\section{Introduction}
Spiral disk galaxies are among the most exquisite and distinctive objects visible in our Universe. They constitute the majority of $L^*$ galaxies in field environments \citep[e.g.,][]{Marzke1998, Tempel2011, Conselice2014}---$L^*$s being the characteristic luminosity systems that dominate both the stellar mass density and the star formation rate density of the Universe at all epochs \citep[e.g.,][]{ Madau2014, Somerville2015}. While we understand how an average $L^*$ galaxy evolves at a population level \citep[e.g.,][]{Behroozi2013, Papovich2015}, two morphologically similar $L^*$ spirals (e.g., Milky Way and M31) can have radically different histories, with distinct secular and external processes competing for dominance \citep[e.g.,][]{Ruiz-Lara2026, Williams2017}.

These galaxies form, grow, and evolve through continued gas accretion, interactions with neighboring galaxies, spiral density waves, internal viscous evolution, and feedback driven by star formation and/or the central supermassive black hole \citep[SMBH; e.g.,][]{Dalcanton1997, vandenBosch2001, Bell2002, Hopkins2014, Pettitt2018, Shu2016, Alexander2025}. As a disk develops through these processes, large-scale radial trends may imprint themselves on the galaxy disk \citep[e.g.,][]{Munoz-Mateos2007, Dale2016}. While secular processes work to blur radial trends over many dynamical timescales \citep[$\sim\,$Gyr; e.g.,][]{Sellwood2002, Kormendy2004, Vera-Ciro2014}, mergers and galaxy-galaxy interactions, can rapidly and substantially impact the distribution of stars and gas \citep[e.g.,][]{Barton2000, Moreno2019}. Disentangling these competing influences in any individual galaxy requires a record of when and where it formed its stars.

That record is the galaxy's star formation history (SFH)---its star formation rate (SFR) as a function of time and, ideally, position. On short timescales ($\lesssim$ a few hundred Myr), it traces ongoing astrophysical processes as they happen---how feedback, gas flows, and dynamical perturbations sculpt star formation across the disk \citep[e.g.,][in a dwarf galaxy context]{Dohm-Palmer2002,Weisz2009, McQuinn2010}. On longer timescales (many Gyr), it instead records an integrated history of when a galaxy experienced epochs of heightened or suppressed star formation. Connecting these to physical drivers typically requires external context such as a record of its merger history from the stellar halo or the SFH of its neighbors, with comparisons to cosmological simulations to guide intuition \citep[e.g.,][]{DSouza2018}. In this work we exploit both timescales, with an emphasis on the former: we measure recent, spatially resolved SFHs to probe the astrophysics of star formation on short timescales, and draw on the ancient SFH, interpreted alongside environmental context and cosmological simulations, to understand what may have driven the major episodes in a galaxy's past.

Reconstructing the history in this way, however, is the exception rather than the rule. Most observational tracers of a galaxy do not recover a SFH at all and instead measure an average recent SFR \citep[usually $\lesssim100\,$Myr; e.g.,][]{Hao2011,K&E2012}. They rely on the integrated light of stellar populations (e.g., UV continuum) or trace the response of the interstellar medium to the light emitted by stellar populations \citep[e.g., H$\alpha$ line and infrared luminosities;][]{Kennicutt1998}. While useful for large galaxy surveys and for generalizing to high redshifts \citep[e.g.,][]{Leroy2019, Madau2014}, these tracers are limited by dust systematics and are insensitive to the burstiness of star formation on $\sim10$--$100\,$Myr timescales as they are typically calibrated to a constant SFH. 

Recovering the SFH, rather than a single recent-average rate, requires a different approach. For nearby galaxies, we can infer spatially resolved SFHs by directly resolving individual stars in different regions, performing photometry, and fitting the observed color-magnitude diagrams (CMDs). This enables us to characterize the distribution of recent star formation, its relationship to earlier star forming episodes, disk-wide asymmetries, and how global SFRs can change from sub-dynamical timescales to longer timescales. Crucially, SFHs measured this way provide an independent benchmark against which the integrated-light SFR indicators above can be tested and calibrated. This form of galactic archaeology has been applied to nearby dwarfs and sub-$L*$ spirals \citep[e.g.,][]{Weisz2008, McQuinn2010, Sacchi2019, Lazzarini2022, Correnti2025} but extending it to larger systems is demanding, and has been accomplished for only two $L^*$ galaxies---M31 and NGC 6946 \citep[e.g.,][]{Lewis2015, Williams2017, Tran2023, Wainer2026}. 

For M31, spatially resolved SFHs spanning several Gyr reveal a galaxy-wide decline in star formation over the last $\sim1.2\,$Gyr, likely in the aftermath of a merger event $2-4\,$Gyr ago. In NGC 6946, UV-based SFHs are constrained to much shorter timescales ($\lesssim25\,$Myr), yet similarly point to a coherent, galaxy-wide suppression of star formation over that period. Whether this shared trend reflects a common mode of evolution among $L^*$ spirals, or whether different environments produce markedly different SFHs as simulations predict \citep[e.g.,][]{deLucia2006, Garrison-Kimmel2019}, cannot be answered from two systems alone. The nearby $L^*$, M81 offers a compelling opportunity to begin addressing such questions.

Located at a distance of $\sim3.8\,$Mpc, M81 is a Milky Way mass SA(s)ab grand design spiral galaxy, that is interacting with two main companions---the starburst galaxy M82 and the disrupted elliptical galaxy NGC 3077. M82's starburst is widely attributed to tidal forcing by M81, though the timing of the interaction is debated: estimates of M82's last pericenter with M81 range from $150\,$Myr ago to $\sim1\,$Gyr ago \citep[e.g.,][]{Brouillet1991, Yun1999, deGrijs2001, Mayya2006, Li2015}. NGC 3077, a visually smooth galaxy with a star-forming core, has been largely stripped of its HI gas. This gas contributes to an extensive tidal debris field---comprising bridges, streamers, and star forming filaments---that reveal the trio's rich interaction history \citep[e.g.,][]{Yun1994, deBlok2018, Okamoto2015, Smercina2020}. M81 is thus an important laboratory for studying the impact of galaxy-galaxy interactions on the disk of a large spiral galaxy.

Large parts of M81's stellar disk were mapped and studied for the fist time by \citet{Davidge2009} using the Canada-France-Hawaii Telescope. Based on the spatial distributions of red supergiant (RSG) and bright main sequence (MS) stars, they found that star formation was curtailed in the central $\sim4\,$kpc of the galaxy in the last $\sim100\,$Myr. \citet{Williams2009} conducted a deep pencil-beam study of stars in M81's outer disk containing part of its western spiral arm using Hubble Space Telescope's (HST) ACS instrument. Leveraging the morphology of the red-clump feature, they derived SFHs up to ancient times and found that most stars in the disk had formed by $z\sim1$ and that the spiral arm had a lifetime of $\gtrsim 100\,$Myr. \citet{Choi2015} conducted a spatially-resolved SFH analysis of M81's disk using ACS. Their study, being limited to M81's western spiral arm and to ages $\lesssim150\,$Myr, found no evidence of star formation propagation across the spiral arm. They thus rule out stationary density waves with a single pattern speed and instead conclude that M81's spiral arms are tidally-induced density waves. 

A spatially-resolved global SFH analysis of M81's disk, however, has never been done. In this paper, we use legacy data and photometry from the ACS Nearby Galaxy Survey Treasury \citep[][]{Dalcanton2009} to derive the SFH of M81 over the last $\sim180\,$Myr. We map M81's young stellar populations well beyond the optical extent of the galaxy disk ($D_{25} \sim 24.2\,$kpc), examine radial trends and disk-wide asymmetries in SFHs, compare our CMD-based SFRs to SFR indicators, compare global SFH trends with M31 and NGC 6946, and compile M81's SFH evidence to constrain its merger history. 

\begin{figure*}[t]
    \centering
    \includegraphics[width=\linewidth]{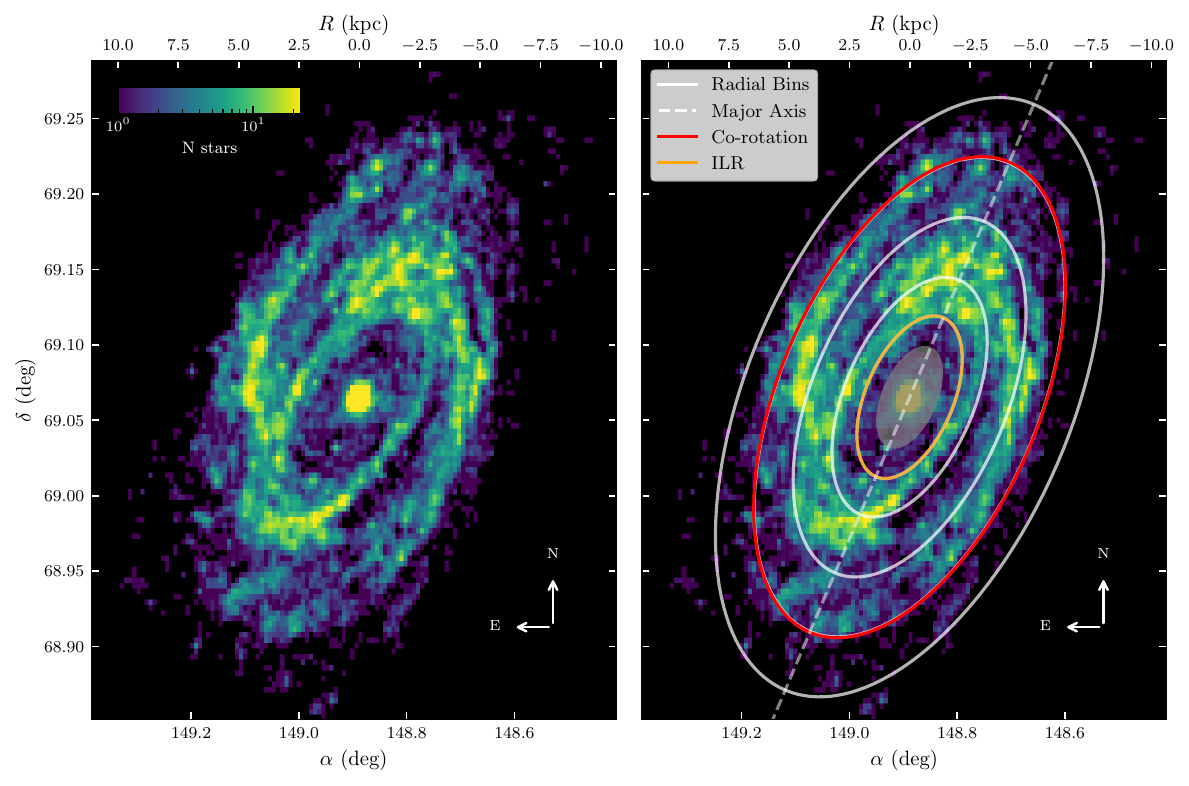}
    \caption{Our M81 survey footprint showing the density map of $\lesssim100\,$Myr stars. The ellipses denote the boundaries of the five radial bins used in our analysis. They are situated at galactic radii of $\sim2.4\,$kpc, $\sim3.8\,$kpc, $\sim5.6\,$kpc, $\sim8.4\,$kpc, $\sim11.2\,$kpc, and $\sim14.0\,$kpc. The central $\sim2.4\,$kpc region (gray patch) has been excluded throughout due to poor photometry. The red ellipse denotes the co-rotation radius at $11.24\,$kpc. Dashed white line represents the major axis of M81. A ring-like feature is seen at $\sim4\,$kpc.}
    \label{fig:spatial-map}
\end{figure*}

The paper is organized as follows. In Section \ref{sec:data}, we describe our resolved star dataset and photometry. In Section \ref{sec:sfh}, we describe our procedure for measuring SFHs. In Section \ref{sec:results}, we describe our spatial maps, examine spatial and temporal trends in SFHs, and present a global SFH. In Section \ref{sec:sfr-indicators}, we compare our CMD-based SFRs to UV-based SFR indicators. Finally, we explore the galaxy-evolution physics governing the SFH trends in Section \ref{sec:physics} and attempt to constrain M81's merger history in Section \ref{sec:merger-history}. Throughout the paper, we assume a distance modulus of $d_{\mathrm{mod}} = 27.9$ \citep{Williams2009} for converting between angular measurements and physical distances.


\section{Data}\label{sec:data}

The M81 resolved-star dataset used in this work is archival and based on the optical imaging of 29 M81 fields using HST's Advanced Camera for Surveys (ACS; Proposal ID: 10584; PI: Zezas). All fields were imaged in at least two filters--- F435W and F606W (a few outer disk fields were additionally imaged in F814W). Together, these 29 fields are contiguous and cover the entire optical extent of the galaxy.

\begin{figure*}
    \centering
    \includegraphics[width =\linewidth]{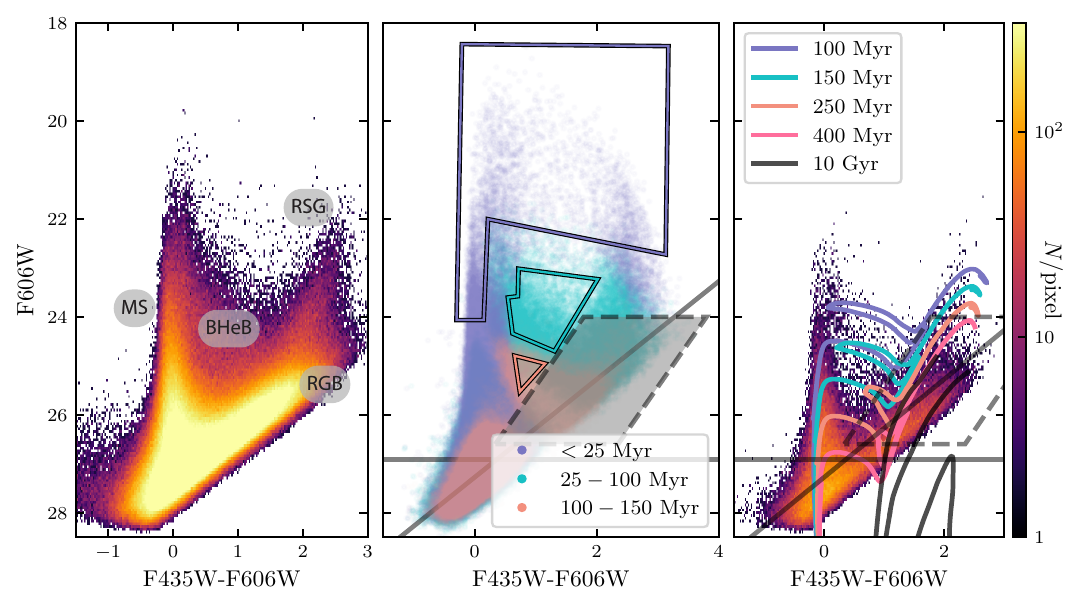}
    \caption{(Left) Hess diagram of all GST sources analyzed in M81. Prominent morphological features have been labeled. (Center) Model CMD of M81 generated using M81's best-fit SFH. $\lesssim25\,$Myr stars are shown in blue, $\sim25-100\,$Myr stars are shown in cyan, and $\sim100-150\,$Myr stars are shown in salmon. Regions used to select stars from each age range from the observed CMD for generating spatial maps are also shown. Gray lines denote the 75\% completeness limits of our outermost radial shell.  (Right) Hess diagram of Shell 5 East--- one of our deepest regions. Representative PARSEC \citep{Bressan2012-PARSEC} isochrones are overplotted to indicate the locations of stars $\lesssim400\,$Myr. RGB stars are represented by $10\,$Gyr isochrones ([M/H]=0.0, -1.0, -2.0). Dashed gray line indicates the exclusion zone used to remove ancient RGB stars and blends. Grey lines indicate the 75\% completeness limits of the two filters.}
    \label{fig:example-cmd}
\end{figure*}

Photometry for all 29 fields was measured as part of the ANGST survey \citep{Dalcanton2009}. In brief, bias and flat-field corrections were performed using the STScI ACS pipeline OPUS. Point spread function (PSF) photometry was obtained using the ANGST pipeline, which is optimized for ACS images through the ACS module within the \texttt{DOLPHOT} photometry package \citep{Dolphin2000PHOT}. Additional details of the photometry procedure can be found in \citet{Dalcanton2009}. Following the work of \citet{Choi2015}, we create the final stellar catalog by choosing sources (GSTs) with \texttt{SNR}$>4$ in both F435W and F606W filters, (\texttt{sharp}$_{\mathrm{F435W}}$ $+$ \texttt{sharp}$_{\mathrm{F606W}})^2 < 0.075$, and (\texttt{crowd}$_{\mathrm{F435W}}$ $+$ \texttt{crowd}$_{\mathrm{F606W}}) < 0.6$. These cuts ensure high quality photometry by selecting point-like sources, which are not significantly affected by crowding. Finally, we merge the stellar catalog of the 29 fields into a unified mosaic. We remove duplicates by choosing stars from a single field in regions where multiple fields overlap and mask the bright foreground Milky Way stars and their diffraction spikes. These selections yield a homogeneous catalog sensitive to stars tracing the last $\sim150\,$Myr of star formation across almost the whole face of M81 (see Figures \ref{fig:spatial-map} and \ref{fig:example-cmd}). We also make use of artificial star tests (ASTs) from both \citet{Dalcanton2009} and \citet{Choi2015}, which sample the full range of stellar densities and depth across the survey footprint.

In the M81 disk, the depth of the photometry in each filter is set by crowding---the 75\% completeness limit increases with increasing galactocentric distance (Figure \ref{fig:shell-cmds}). Even with the resolution of HST, faint stars in the galactic disk that would otherwise be detected in sparse fields remain unresolved in the dense field of brighter stars. Additionally, many faint stars are biased brighter by blending with neighboring stars. These blended sources are especially a concern at depths where the magnitude limit lies near a rapid increase in the luminosity function of the underlying stellar population. Because faint stars vastly outnumber bright ones, photometric scatter near the detection limit preferentially pushes faint stars to brighter magnitudes \citep[Eddington bias;][]{Eddington1913, Hogg1998}, and these can outnumber the stars with reliable photometry. In our case, the F435W magnitude limit lies close to the tip of the red giant branch (TRGB), where there is a rapid increase in the luminosity function towards fainter magnitudes due to the ancient, several Gyr, red giant branch (RGB) stars (see the example F435W-F606W CMD, third panel Figure\,\ref{fig:example-cmd}). 

The ASTs give us a fairly good sense of the completeness of the photometry as a function of magnitude and the magnitude of photometric errors. Spurious blends typically dominate within $\sim1-2\,$mag of the magnitude limit of each filter. Thus, we only use sources brighter than the $75\%$ completeness limit of each filter for obtaining SFHs. For F435W and F606W, the $75\%$ completeness limits are $\sim1$ and $\sim1.5\,$ mag brighter than the faintest GSTs, respectively.

The HST data and ancillary GALEX, and Spitzer data presented in this article were obtained from the Mikulski Archive for Space Telescopes (MAST) at the Space Telescope Science Institute and the Spitzer Local Volume Legacy (LVL) made available by the Infrared Science Archive (IRSA) at IPAC. The specific observations analyzed can be accessed via \dataset[MAST DOI: 10.17909/q34b-bk25]{https://doi.org/10.17909/q34b-bk25} and \dataset[IRSA DOI: 10.26131/IRSA414]{10.26131/IRSA414}).

\section{Star Formation Histories}\label{sec:sfh}
In this section, we describe our choice of regions for inferring large-scale trends in M81's SFH, how the CMDs of these regions were fitted with stellar evolution models, and how uncertainties on those fits were estimated.
\subsection{Region Selection}\label{sec:region-selections}

For subdividing the M81 disk, we choose to aggregate stars over wide radial bins to build the statistical power needed to constrain the SFH at each location \citep[see for e.g.,][for a similar analysis of NGC 7793]{Sacchi2019}. This approach is better suited to our science goals than subdividing the disk into small spatial pixels, whose SFH accuracy is sensitive to the patchiness of star formation and dust on small scales \citep[e.g.,][]{Lewis2015, Lazzarini2022}.

We split the stellar disk into five radial bins bounded by galactic radii $\sim2.4\,$kpc, $\sim3.8\,$kpc, $\sim5.6\,$kpc, $\sim8.4\,$kpc, $\sim11.2\,$kpc, and $\sim14.0\,$kpc (see Figure \ref{fig:spatial-map}). On the sky plane, these radial bins are bounded by ellipses whose orientation and minor axes are defined by the position angle of M81's major axis \citep[PA $\sim 157\degr$;][]{Appleton1981} and disk inclination \citep[$i=59\degr$;][]{deBlok2008}, respectively. Our choice of radial bins was influenced both by the work of \citet{Choi2015} and the distribution of neutral hydrogen \citep[H{\sc i};][]{deBlok2018}. The outer bin edges spaced $\sim2.8\,$kpc apart from $\sim5.6-14.0\,$kpc are similar to those used by \citet{Choi2015}, with the bin edge at $\sim11.2\,$kpc lying close to the reported average co-rotation radius $R_{cr}\sim11.24\,$kpc. The inner bin edge at $\sim3.8\,$kpc corresponds to the ``hole" in the H{\sc i} distribution within which there is virtually no observed 21\,cm emission. This radius also corresponds to the inner Lindblad resonance (ILR), which separates the bulge and inner disk from the spiral arms and outer disk \citep[][]{Kaufman1989, Reichen1994}. We choose to exclude the innermost $\sim2.4\,$kpc bulge-dominated region of M81 from our resolved star analysis due to photometric errors, biases, and incompleteness from high crowding.

\begin{figure*}[ht]
    \centering
    \includegraphics[width =\linewidth]{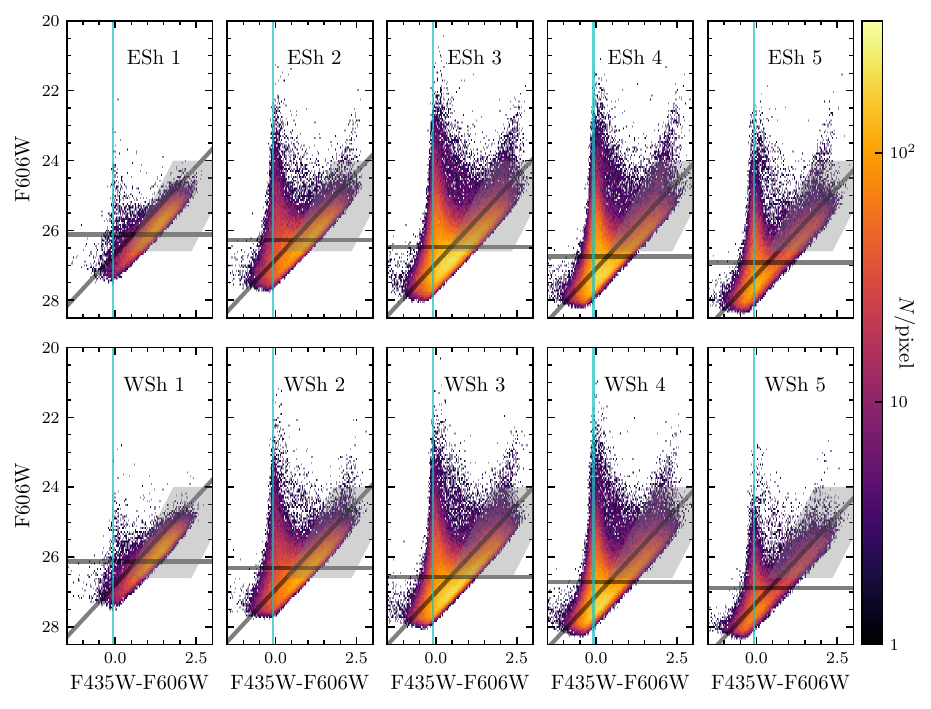}
    \caption{Hess diagrams of all our M81 regions with their 75\% completeness limits. The vertical cyan line represents the color of the un-reddened main sequence. The transparent gray patch shows our CMD exclusion zone. The eastern shells are shown on top and the western shells are shown at the bottom. Galactocentric distances increase left to right. The completeness limits get deeper as we move outward.}
    \label{fig:shell-cmds}
\end{figure*}

The M81 disk is inclined such that the western half is closer to us compared to the eastern half \citep{buta2007devaucouleurs}. Depending on how the young stellar populations are distributed relative to the dust in the ISM, being on the near side or the far side may result in different reddened fractions \citep[e.g.,][]{Dalcanton2023}. Thus, to accurately capture possible systematic differences in dust extinction, we further split M81 along the major axis into an eastern half and western half for our SFH analysis, resulting in ten regions. The F435W-F606W color magnitude diagram (CMD) of each region is shown in Figure \ref{fig:shell-cmds} along with the $75\%$ completeness limit of each filter determined from ASTs. These completeness limits are a strong function of radius, due to less sensitivity to fainter stars in regions of higher stellar density. Additional parameters of each region are listed Table \ref{tab:regions}.

\begin{deluxetable}{lcccccccc}
\tablecaption{Regions of the M81 Disk}\label{tab:regions}
\tablehead{
\colhead{Field} &
\colhead{Semi-major} &
\colhead{F435W} &
\colhead{F606W} &
\colhead{GSTs} &
\colhead{$A_V$} &
\colhead{$dA_V$} &
\colhead{Shell Area} &
\colhead{GSTs}\\
\colhead{} &
\colhead{Axis (kpc)} &
\colhead{(75\%)} &
\colhead{(75\%)} &
\colhead{} &
\colhead{} &
\colhead{} &
\colhead{(kpc$^2$)} &
\colhead{/kpc$^2$}
}
\startdata
Shell\,1 East & 3.1 & 26.67 & 26.12 & 12744 & 0.266 & 1.2 & 6.94 & 1835.63\\
Shell\,1 West & 3.1 & 26.79 & 26.12 & 12843 & 0.266 & 1.1 & 6.94 & 1849.89\\
\hline
Shell\,2 East & 4.7 & 26.83 & 26.28 & 67108 & 0.266 & 1.2 & 13.53 & 4958.76\\
Shell\,2 West & 4.7 & 26.93 & 26.31 & 69860 & 0.266 & 1.2 & 13.53 & 5162.11\\
\hline
Shell\,3 East & 7.0 & 26.95 & 26.48 & 196607 & 0.266 & 1.1 & 31.35 & 6270.64\\
Shell\,3 West & 7.0 & 27.02 & 26.57 & 160177 & 0.266 & 0.9 & 31.35 & 5108.73\\
\hline
Shell\,4 East & 9.8 & 27.10 & 26.75 & 189387 & 0.266 & 1.0 & 43.90 & 4314.54\\
Shell\,4 West & 9.8 & 27.11 & 26.71 & 170259 & 0.266 & 0.9 & 43.90 & 3878.78\\
\hline
Shell\,5 East & 12.6 & 27.28 & 26.92 & 100002 & 0.266 & 0.7 & 56.44 & 1771.94\\
Shell\,5 West & 12.6 & 27.31 & 26.90 & 55209 & 0.266 & 0.7 & 56.44 & 978.25\\
\enddata

\tablecomments{The semi-major axes are the central values of each elliptical bin and represent the physical distance from the galaxy center. The number of GSTs is the number of stars used to fit each region after excluding the RGB. $A_V$ and $dA_V$ represent the best-fit values from the CMD fits. Shell areas are projected values.}
\end{deluxetable}

\subsection{Deriving SFHs}
The M81 photometry lacks the depth required to resolve older ($\gtrsim 1\,$Gyr) main-sequence turnoffs or the red clump, and therefore cannot accurately constrain the ages of the oldest stars \citep[e.g.,][]{Williams2017, Weisz2014, Girardi2016}. Additionally, since our focus is only on the young stellar populations tracing the recent SFH, we exclude from our analysis the CMD region occupied by the ancient RGB stars (Figure\,\ref{fig:example-cmd}). Our conservative CMD exclusion zone, which extends $\sim1\,$mag brighter than the brightest $[M/H]=-2$ ancient RGB isochrone, further removes blended sources. The regions of the CMD that remain for our SFH analysis contain stars with ages that are at most $\sim400$\,Myr.

We derived the SFH of each M81 region using the CMD fitting code \texttt{MATCH} \citep{Dolphin2002, Dolphin2012, Dolphin2013}. \texttt{MATCH} determines the set of single-burst stellar population CMDs that best reproduces the observed CMD after incorporating photometric biases, completeness from ASTs, and dust extinction. The best-fit SFH is then established using a Poisson maximum likelihood technique, which compares the star counts in each CMD bin between the data and the model. The $\chi^2$ equivalent in this case is captured by a fit parameter, which is defined as $-2\ln$ of the Poisson likelihood ratio, $2\sum_i \left[m_i - n_i + n_i\ln(n_i/m_i)\right]$, where $n_i$ and $m_i$ are the observed and model counts in CMD bin $i$ and lower values indicate better fits \citep[see][for more details]{Dolphin2002, Dolphin2012, Dolphin2013}.

Our procedure for obtaining recent SFHs from these restricted CMDs is similar to that used by \citet{Lewis2015}, \citet{Lazzarini2022}, and \citet{Rao2025}. Essentially, the SFRs determined by \texttt{MATCH} hinges on the morphology of the main sequence (MS) stars and the red and blue core helium burning stars (BHeB and RHeB, respectively). The core helium burners in particular serve as robust chronometers for recent star formation episodes, as their distinct colors and magnitudes allow precise age determinations \citep{Dohm-Palmer2002, Weisz2008, McQuinn2012}.

The input parameters for our SFH fits were: distance modulus, choice of stellar evolution model, IMF, extinction, metallicity, binary fraction, magnitude limits, and age bins. We fixed the distance modulus ($m-M$) to 27.9 \citep{Williams2009, Choi2015}, set the magnitude limits to 75\% completeness from ASTs, fixed the binary fraction at 0.35, adopted a \citet{Kroupa2001} IMF, and chose the PARSEC stellar evolution models \citep{Bressan2012-PARSEC, Chen2014-PARSEC, Chen2015-PARSEC, Fu2018-PARSEC}. We used logarithmic age bins from 6.6 to 8.7 (approximately $4–500\,$Myr) with a spacing of 0.15 dex and an additional $9.0-9.5$ age bin (approximately $1-3\,$Gyr) to account for potential photometric errors from old populations. We enforced a monotonically increasing age-metallicity relationship over our age range using the \texttt{zinc} option to capture enrichment from star formation and to avoid nonphysical metallicities --- a recommended strategy. Even though we used age bins up to $\sim3\,$Gyr to fit the CMD, we remind the reader that we are only sensitive to stars up to $\sim150\,$Myr in age and hence report our SFHs up to the time bin that straddles this approximate limit ($\log\,t=8.1-8.25$ or $\sim125-180\,$Myr). We obtain SFHs of composite regions (radial shells, half disk, and full disk) by combining the SFHs of each half-shell region using the \texttt{outcombine} functionality of \texttt{MATCH}.

We accounted for extinction arising from both foreground dust and the dust local to M81. We parameterize foreground extinction by $A_V$, which reddens all stars uniformly. We capture variable extinction from the patchy distribution of dust in the spiral arms and the inter-arm regions through the differential extinction parameter $dA_V$. This simple model is well-suited for young populations (age $<1$\,Gyr), which typically experience a top-hat–like distribution of reddening \citep[e.g.,][]{Dolphin2003, Weisz2014, Lazzarini2022}. For the synthetic CMDs that it generates, \texttt{MATCH} applies extinction to all stars in a uniform distribution between $A_V$ and $A_V + dA_V$. Thus, in CMD space, $A_V$ shifts the entire population fainter and red-ward, whereas $dA_V$ broadens and dims features like the main sequence (MS). We fixed $A_V$ to 0.266\,mag \citep{Schlegel1998} as we don't expect the foreground extinction to vary by much over the face of the disk and determined the best values for $dA_V$ by minimizing the fit parameter obtained by running \texttt{MATCH} multiple times over a range of $dA_V$ values. We initially did a coarse search between 0 and 1.4 mag, with steps of 0.2 mag; since the fits tended to prefer $dA_V$ values on the high side, we refined the search range between $dA_V$ of 0.7 and 1.3 mag, with steps of 0.1 mag.

\begin{figure}
    \centering
    \includegraphics[width=\linewidth]{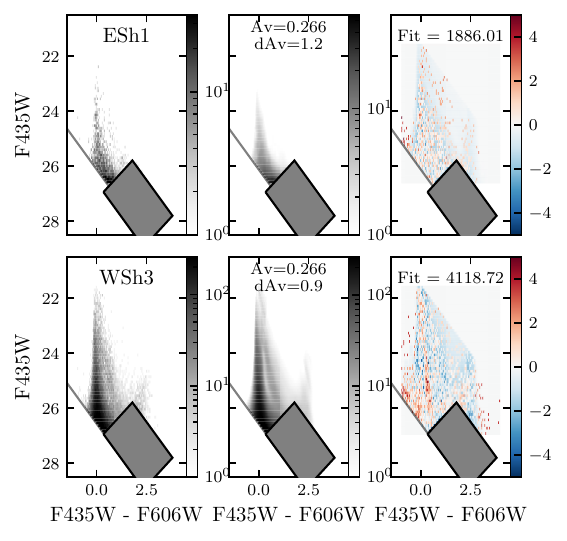}
    \caption{Model Hess diagrams and their quality of fit for two representative regions--- Shell\,1 East (Top) and Shell\,3 West (Bottom)--- as output by MATCH with the short wavelength on the y-axis. (Left) Observed Hess diagram used by \texttt{MATCH} that balances number statistics on the CMD with time resolution in the SFH; (Center) best-fit model Hess diagram using PARSEC stellar evolution models; (Right) residual significance Hess diagram. The grayscale in the left and center panels represent the number of stars per CMD bin. In the right panel, the color bar scaling reflects the significance of each pixel in the residual (data-model) relative to the standard deviation of a Poisson distribution. The RGB masks used are shown in gray in all panels along with the F606W 75\% completeness limits.}
    \label{fig:cmd-fits}
\end{figure}

We show the quality of our Hess diagram fits using PARSEC models for two representative regions---Shell\,1 East and Shell\,3 West---in Figure \ref{fig:cmd-fits}. The scaling of the residual Hess diagram reflects the pixel-wise deviation of the observed Hess diagram from the model relative to the standard deviation of a Poisson distribution. Overall, the models reproduce distinct features of the observed Hess diagrams such as the horizontal branch, red supergiant sequence, and the main sequence, matching the data well in terms of densities of stars, luminosities, colors, and scatter. Some residual mismatch between the data and best-fit model CMDs remains, most noticeably in the bright, sparsely populated regions occupied by blue and red supergiants. The models of these phases are sensitive to the adopted mass-loss prescriptions, metallicity, and high-mass IMF, and such residuals are common in CMD fitting of young stellar populations. In the following section, we describe how our systematic uncertainties account for these model shortcomings. We show and discuss trends in best-fit metallicities in Appendix\,\ref{sec:metallicities}.

\subsection{SFH Uncertainties}
Uncertainties in our SFH analysis arise from both random and systematic uncertainties. Random uncertainties stem from the finite number of stars, and therefore generally scale inversely with this number. We calculated random uncertainties using a hybrid Monte Carlo(MC) process \citep{Duane1987} implemented through \texttt{MATCH} \citep[see][for implementation details]{Dolphin2013}. The result of this Markov Chain MC routine is a sample of 10,000 SFHs whose density in parameter space is proportional to its probability density. Upper and lower random error bars are calculated by identifying the boundaries of the region containing 68\% of the samples. The advantage of this method is that it accounts for cases where the most likely SFH has zero star formation in individual time bins, allowing the uncertainties on the star formation rate in such bins to be estimated.

Systematic uncertainties stem from inherent shortcomings of the stellar evolution model used to measure the SFHs (e.g., uncertainties in mass loss, convective overshoot, etc.). We were unable to characterize these uncertainties by comparing our PARSEC models with other stellar evolution libraries such as BASTI18 and MIST, since we found that those models do not adequately cover the full range of stellar ages needed to accurately reproduce the youngest and brightest regions of the CMD. Instead, we adopted an alternative approach that assesses the magnitude of systematic errors by randomly shifting the PARSEC model CMDs and re-measuring SFHs. The random shift method was calibrated against differences in multiple model suites, and has been shown to be more conservative than, if not consistent with, comparing against any two chosen evolutionary libraries \citep[e.g.,][]{Skillman2017, Smercina2025}.

We start by obtaining 50 Monte Carlo SFH realizations of the best fit model CMD. For each realization, we shift the model CMD in $\log(T_{eff})$ and $M_{bol}$ and remeasure the SFH following the procedure described in \citet{Dolphin2012}. The shifts are randomly chosen from a Gaussian distribution with $\sigma_{\log T_{eff}} = 0.012$ and $\sigma_{M_{bol}} = 0.18$. From the resulting 50 SFHs, we determine the upper and lower systematic error bars by calculating the range that contains 68\% of the samples. As discussed in \citet{Dolphin2012}, this technique produces 1$\sigma$ systematic uncertainties that include the best fit SFH as derived using the PARSEC stellar evolution models. 

Notably, \citet{Smercina2025} also demonstrate that uncertainties estimated using such a method are sufficient to capture uncertainties up to $1.5\,$mag in distance modulus and adopted $A_V$, as shifts from these parameters in CMD space mirror differences in stellar atmospheres when using different evolutionary model suites. We confirm this directly for M81: re-measuring the SFH with the distance modulus fixed to $27.8$ \citep[the value reported by][]{Radburn-Smith2011}, rather than our adopted value, yields an SFH that lies within our systematic error bars. Because the differences between models, distance, and reddening are highly covariant, this approach allows us to capture these degenerate sources of uncertainty in a non-parametric way.

\section{Results from Resolved Stars}\label{sec:results}
In this Section, we describe the spatial maps of different stellar populations in M81, discuss radial trends in the SFH, compare the SFHs of two halves of the M81 disk, and finally widen our lens to look at the global SFH within our survey footprint.

\subsection{Spatial Maps}
\begin{figure*}
    \centering
    \includegraphics[width=0.9\linewidth]{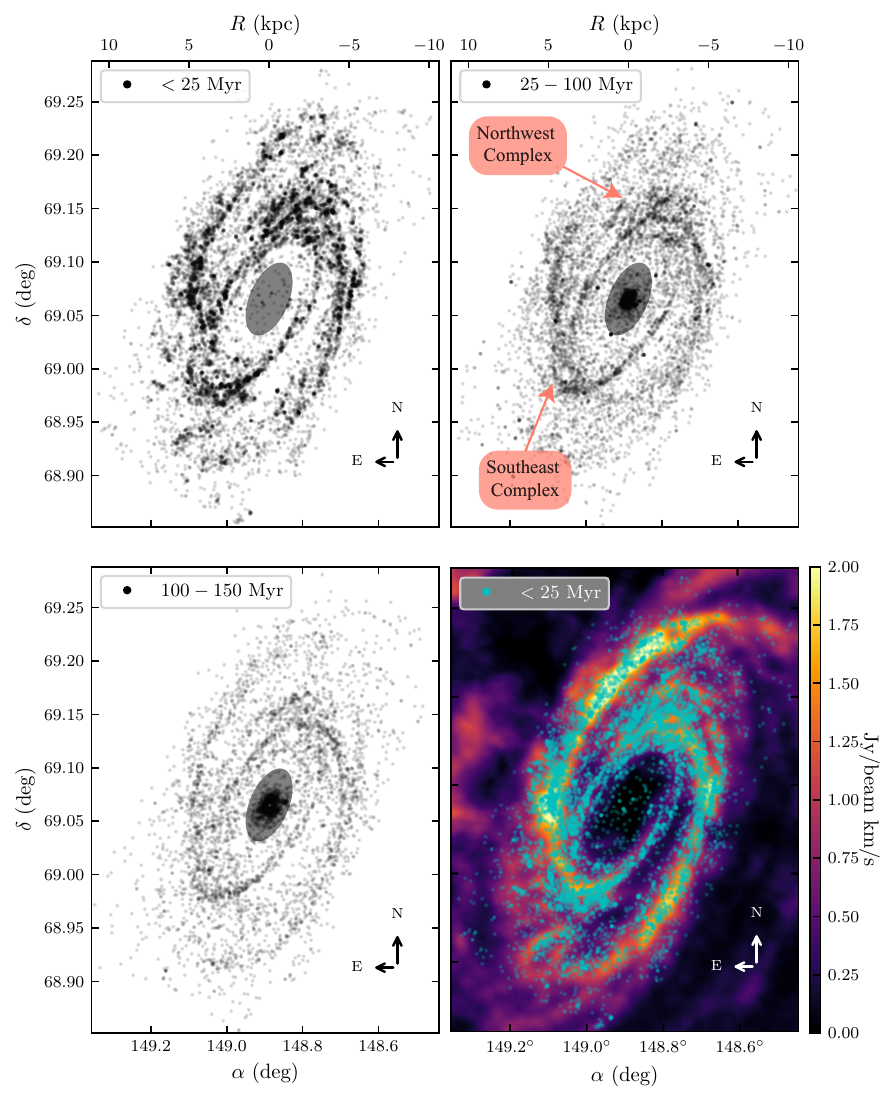}
    \caption{(Top Left) Map of $\lesssim25\,$Myr stars. (Top Right) Map of $\sim25-100\,$Myr stars with the northwest and southeast star complexes labelled. (Bottom Left) Map of $\sim100-150\,$Myr stars. The innermost $\sim2.4\,$kpc has been grayed out as this region was excluded from our analysis due to crowding. (Bottom Right) Map of $\lesssim25\,$Myr stars superimposed on the map of neutral hydrogen \citep{deBlok2008}. These stars closely follow the gas distribution. }
    \label{fig:age-maps}
\end{figure*}

The spatial distribution mosaics of stars in different age ranges show interesting features (Figure \ref{fig:age-maps}). The stars in each age range ($<25\,$Myr, $25-100\,$Myr, and $>100\,$Myr) were selected by identifying unique regions of the CMD that are populated by corresponding model stellar populations. We generated these model populations using global best-fit SFRs, our adopted $A_V$, and the weighted average of best-fit $dA_V$ from our ten regions (center panel, Figure \ref{fig:example-cmd}). The model CMD also justifies our choice to present SFHs only for lookback times $\lesssim180\,$Myr our oldest time bin is centered at $\sim150\,$Myr). The CMD regions used to constrain the SFH---occupied by upper main sequence and core helium burning stars---are only appreciably populated by stars younger than $\sim150\,$Myr.

The $<25\,$Myr map shows two well defined spiral arms (henceforth, eastern and western spiral arms) with stars closely following the distribution of neutral hydrogen. Large complexes of stars can be seen at the northern side of the western spiral arm from $\sim3.8-8.4\,$kpc, and the southern and eastern sides of the eastern spiral arm from $\sim5.6-11.2\,$kpc. We will henceforth refer to them as the northwest complex and southeast complex, respectively. These complexes are also prominent in the maps of $25-100\,$Myr stars and the $>100\,$Myr stars, despite the spiral patterns being less pronounced. Notably, the definition of the spiral arms at larger radii decreases with increasing age of the population. 

In addition to the spiral features, the number of $<25\,$Myr stars also drastically decreases $\lesssim3.8\,$kpc compared to larger radii. This is further reflected in the H{\sc i} map, which shows almost no emission $\lesssim3.8\,$kpc. However, the relative dearth of $<25\,$Myr stars in the innermost shell is not correspondingly observed in the $25-100\,$Myr or the $>100\,$Myr age ranges.

\subsection{Radial Star Formation Rate Densities}
\begin{figure*}
    \centering
    \includegraphics[width=0.9\linewidth]{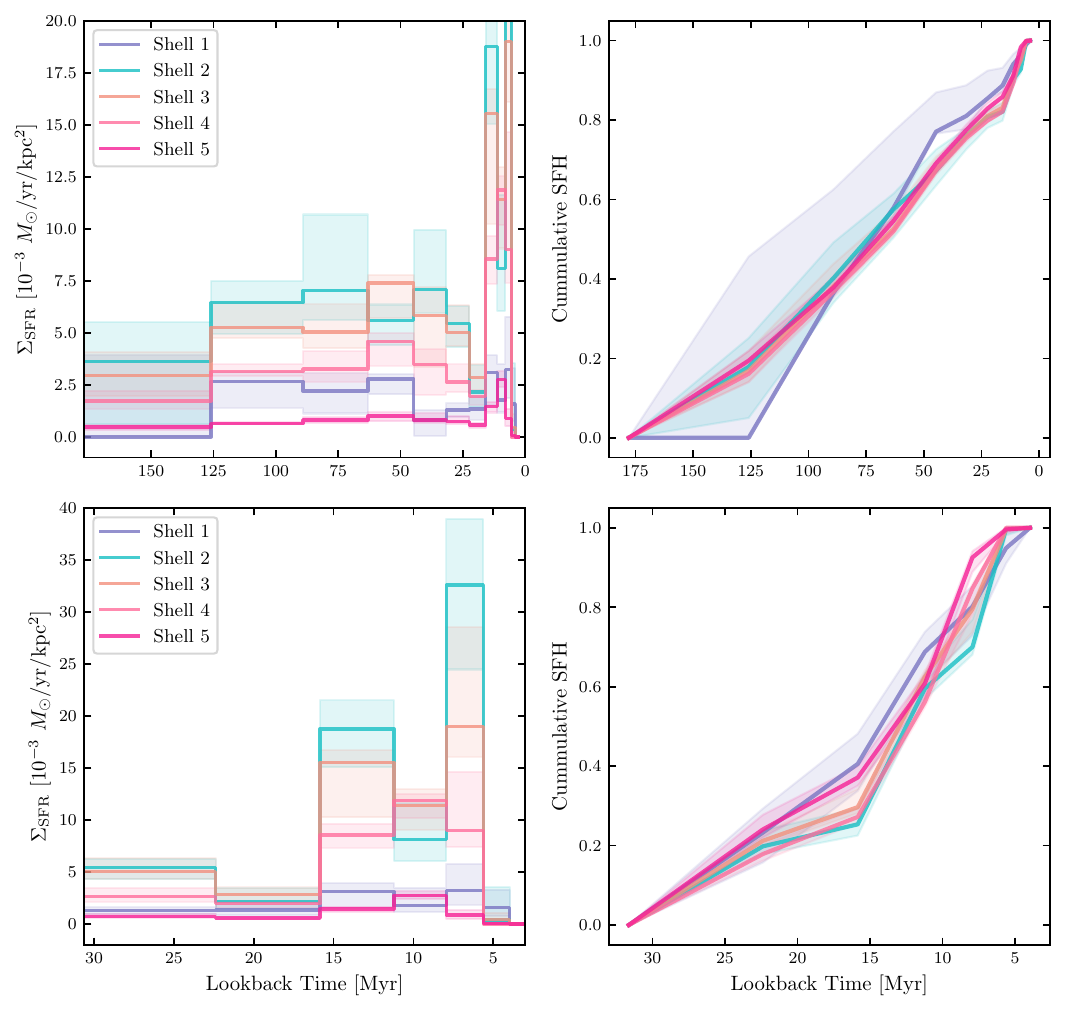}
    \caption{(Top Left) Star formation rate densities (per projected area) as function of lookback time for each of our radial bins $\sim4-180\,$Myr. (Top Right) Cumulative fraction of stars formed in each of our radial bins as a function of lookback time since $\sim180\,$Myr. (Bottom Left) Star formation rate densities $\sim4-30\,$Myr. (Bottom Right) Cumulative fraction of stars formed since $\sim30\,$Myr. Error envelopes represent the combination of random and systematic uncertainties.}
    \label{fig:sfh}
\end{figure*}

In Figure \ref{fig:sfh}, we show the SFR surface densities of each radial shell (eastern and western halves combined), using projected shell areas. We also show the cumulative mass fraction of stars formed prior to $\log\,t=8.25$ ($\sim180\,$Myr) and $\log\,t=7.5$ ($\sim32\,$Myr) in each radial shell. The SFRs are mostly constant over the $\sim180\,$Myr range and increase slightly $\lesssim15\,$Myr. Shell 1 SFR densities are lower throughout compared to Shell 2. Barring this, SFR densities increase with decreasing radius (see also Figure \ref{fig:cmd-hybrid-comparison}). Additionally, Shell 1 SFRs appear to ramp up $\sim120\,$Myr ago and slow down $50-25\,$Myr ago. This effect is apparent in the cumulative plot (top right panel, Figure \ref{fig:sfh}), where the slope of the Shell 1 SFH steepens in the $\sim90-125\,$Myr time bin and becomes shallower $\lesssim50\,$Myr. Shells 2, 3, and 4, which accommodate the northwest and southeast complexes, also show enhanced star formation $<15\,$Myr as the slopes of their cumulative SFH steepen in that age range. 

\subsection{Disk Asymmetries}\label{sec:results-asymmetries}
\begin{figure}
    \centering
    \includegraphics[width=\linewidth]{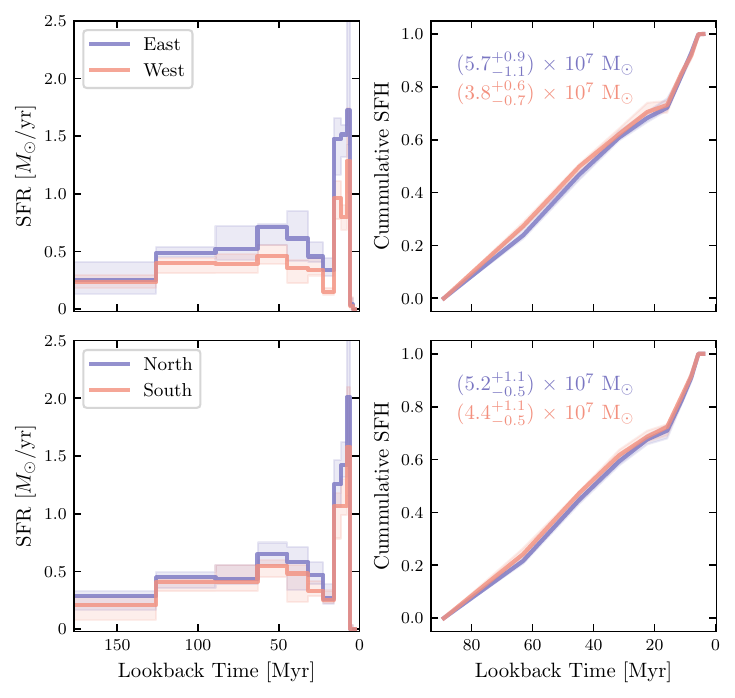}
    \caption{(Top Left) SFR of the eastern half (blue) and western half (salmon) of the M81 disk as a function of lookback time. (Top Right) Cumulative SFHs of the eastern and western halves and the total stellar mass formed in each half disk over the last $\sim90\,$Myr. The total stellar mass formed in the eastern half is $\sim50\%$ higher than the western half in this time period. (Bottom Left) SFR of the northern half( blue) and southern half (salmon) of the M81 disk. (Bottom Right) Cumulative SFHs of the northern and southern halves and the total stellar mass formed in each half disk over the last $\sim90\,$Myr. The total stellar mass formed in the northern half is $\sim18\%$ higher than the western half in this time period but, the difference is smaller than the uncertainties.}
    \label{fig:half-disk-sfh}
\end{figure}
The recent SFH of the M81 disk is asymmetric (Figure \ref{fig:half-disk-sfh}). The eastern half of the M81 disk had systematically higher SFRs ($>1\sigma$) in the last $\sim90\,$Myr compared to the western half. However, we find no significant SFR asymmetry between the northern and southern halves upon repeating our SFH analysis independently for these regions. The asymmetry is most apparent in the total stellar mass formed in the last $\sim90\,$Myr, reported in the cumulative plots. Over the last $\sim90\,$Myr, the eastern half formed $\sim$50\% more stellar mass than the western half, a difference that exceeds the systematic uncertainties. The northern half formed only $\sim$18\% more than the southern half, which is within systematic uncertainties. From Table \ref{tab:regions}, we can see that the east-west asymmetry in SFRs arises from fewer GSTs used for SFH fits in western shells 3, 4, and 5 compared to the corresponding eastern shells. These trends also agree with the spatial maps (Figures~\ref{fig:spatial-map}, \ref{fig:age-maps}), where we see that the stars $\lesssim100\,$Myr that follow the eastern spiral arm appear both more dominant and extend to higher galactocentric radii compared to stars of similar age in the western spiral arm. The east-west asymmetry largely disappears in the $\sim90-180\,$Myr age range as differences in SFRs are within $1\sigma$ uncertainties. This agreement is reflected in the symmetric map of $\sim100-150\,$Myr stars in Figure~\ref{fig:age-maps}.

In addition to star formation asymmetries, we find that stars in the western half (near side) experience somewhat lower differential extinction from dust internal to M81 compared to the eastern half (far side; refer Table \ref{tab:regions}). The difference, however, is mostly marginal and comparable to the step size with which we vary $dA_V$.

\subsection{Global SFR}
\begin{figure}
    \centering
    \includegraphics[width=\linewidth]{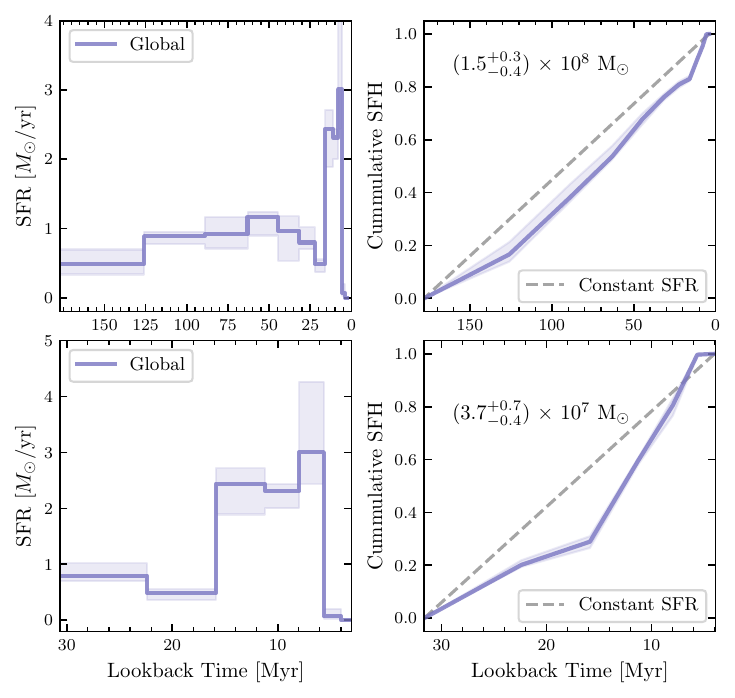}
    \caption{Global SFHs. (Top Left) SFRs of the M81 disk between galactic radii $\sim2.4\,$kpc and $\sim14.0\,$kpc over the last $\sim180\,$Myr. (Top Right) Cumulative SFH over the last $\sim180\,$Myr. Total stellar mass formed in this time span is also reported. The dashed diagonal line corresponds to a constant SFR. (Bottom Left) SFRs of the M81 disk over the last $\sim32\,$Myr. (Bottom Right) Cumulative SFH over the last $\sim32\,$Myr.}
    \label{fig:global-sfh}
\end{figure}

The global SFH of the M81 disk--- obtained by combining the SFHs of all analysis regions--- is approximately constant, if not slightly ramping up, over the last $\sim180\,$Myr (Figure \ref{fig:global-sfh}). This behaviour is apparent from the cumulative history, which runs parallel to the line of constant star formation for most of the time range before steepening in the last $\lesssim15\,$Myr. This uptick in SFRs at younger ages is largely driven by the star formation in Shells 2, 3, and 4. While the SFR enhancement at recent times appears significant from the SFR vs. time plot, we caution the reader from interpreting this too seriously as the enhancement is only modest in terms of stellar mass formed---seen as a modest deviation from the line of constant SFR. We find the global $\sim90\,$Myr mean SFR to be $1.06^{+0.18}_{-0.20}\,M_\odot\,\mathrm{yr}^{-1}$, where the uncertainties include random and systematic uncertainties on the stellar mass formed.

It is worth noting that the global SFH we derive is specifically the SFH of the M81 disk between galactic radii $\sim2.4-14.0\,$kpc, and excludes the inner bulge of the galaxy. Based on our radial trends and the FUV map (Figure \ref{fig:multiwav-map}), we expect the inner bulge to contribute a negligible fraction to this SFH. Additionally, our survey region goes beyond the optical size of the galaxy. The $D_{25}$ ellipse of the galaxy lies at a galactic radius of $\sim12.1\,$kpc (HyperLeda), which makes our survey region $1.30$ times larger than the sky area contained within the $D_{25}$ ellipse. Thus, we expect our global SFH estimate to account for nearly all stars formed in M81's disk in the last $\sim180\,$Myr.

\section{Comparisons with Star Formation Tracers}\label{sec:sfr-indicators}
\begin{figure*}
    \centering
    \includegraphics[width=0.9\linewidth]{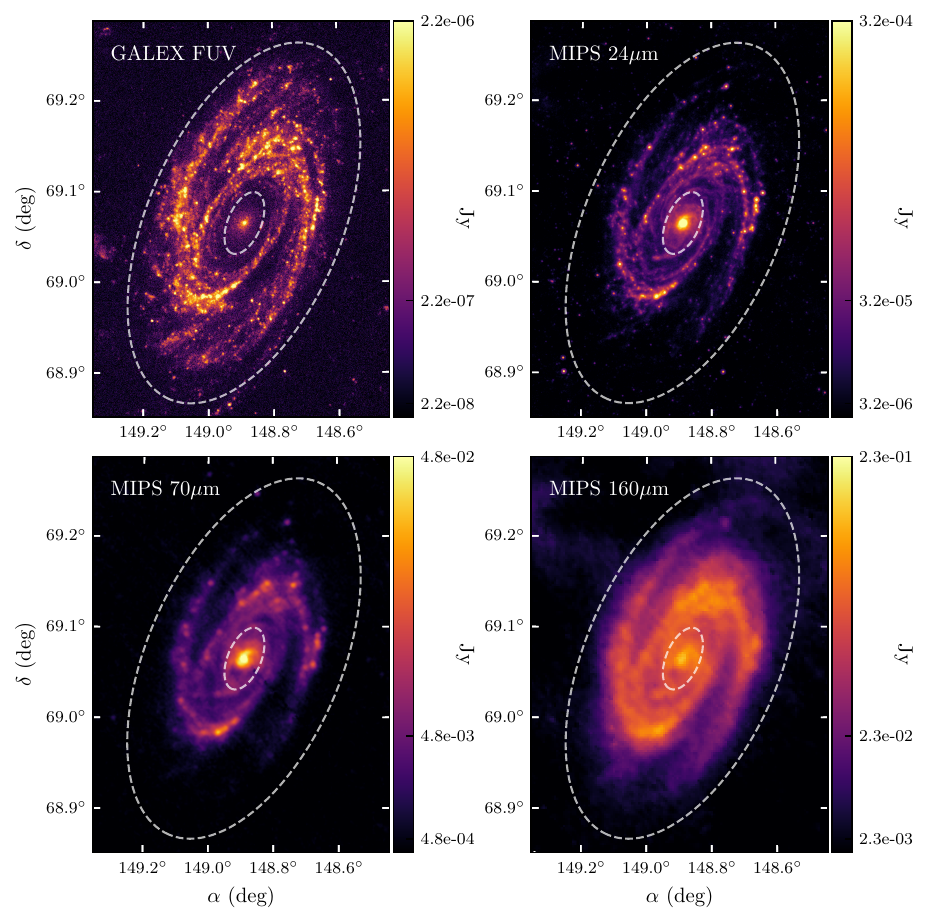}
    \caption{Multiwavelength maps of M81. (Top Left) GALEX FUV map from the Nearby Galaxy Survey. Spitzer Local Volume Legacy \citep{Dale2009} 24$\mu$m (Top Right), 70$\mu$m (Bottom Left), and 160$\mu$m (Bottom Right) maps. Dashed ellipses correspond to galactic radii of $\sim2.4\,$kpc and $\sim14\,$kpc and mark the boundaries of the region used for measuring SFRs in this work.}
    \label{fig:multiwav-map}
\end{figure*}
In this section, we compare our CMD-based SFRs to those obtained using SFR-indicators.

The SFRs of galaxies can be indirectly estimated in a number of ways from integrated light and recombination line observations. These include H$\alpha$ luminosities, UV continuum, IR luminosities, non-thermal radio synchrotron, and X-rays \citep[e.g.,][]{Kennicutt1998,K&E2012}. The UV continuum directly traces the photospheric emission by young stars of ages $\lesssim100\,$Myr and is hence the most relevant tracer to compare our CMD SFRs against. H$\alpha$ luminosities on the other hand capture the response of the ambient ISM to the ionizing radiation emitted by the photospheres of young $\lesssim10\,$Myr stars. Hence, the SFRs calibrated to H$\alpha$ luminosities suffer much greater stochasticity due to variations in ISM properties. However, both these tracers are limited by severe dust attenuation making it challenging to account for all the stars in a galaxy. 

Hybrid SFR prescriptions, which combine UV+IR luminosities (and H$\alpha$+IR for that matter) offer a way around the dust attenuation problem. According to energy budget arguments, all starlight absorbed at UV and optical wavelength by the interstellar dust is re-emitted in the IR. Thus, the combination of UV and IR luminosities should be able to probe the intrinsic UV luminosity. Based on this principle, we obtain SFRs from dust-corrected FUV luminosities with corrections made with 24$\mu$m and total IR (TIR; $3-1100\,\mu$m) luminosities using calibrations from \citet{Hao2011}. 24$\mu$m emission predominantly traces warmer dust associated with intense radiation fields in compact star-forming regions \citep[e.g.,][]{Calzetti2007, Perez-Gonzalez2006}, whereas TIR emission is sensitive to a broader range of dust temperatures, including cooler diffuse dust heated by more distributed older stellar populations. Consequently, TIR-based SFRs may overestimate recent star formation when interpreted solely as reprocessed UV emission from young stars \citep[e.g.,][]{Kennicutt1998, Kennicutt2009}.

The relation between the SFR and FUV luminosity is given by
\begin{equation}
    \mathrm{SFR}\,(M_\odot\,\mathrm{yr}^{-1}) = 10^{-43.35} L_{\mathrm{FUV}}\, (\mathrm{erg\,s}^{-1})
\end{equation}
This calibration was obtained assuming a constant SFH, a \citet{Kroupa2003} IMF with mass limits 0.1-100\,$M_\odot$, and a solar metallicity. Since \texttt{MATCH} reports SFRs by integrating the \citet{Kroupa2001} IMF from $0$ to infinity, we apply a correction factor of $-0.12\,$dex to our CMD SFRs \citep{Telford2020} to account for the finite IMF mass limits and facilitate comparisons with the FUV tracer. This correction does not account for the different high-mass slopes of the two IMFs. For a constant SFH, SFRs based on the \citet{Kroupa2003} IMF must be increased by $\sim13\%$ to be comparable with those based on the \citet{Kroupa2001} IMF. Because this factor depends on the SFH, and M81's recent SFH is not constant, we do not apply it here; quantifying its SFH dependence is beyond the scope of this paper. We discuss its effect on our comparison below.

We show the Galaxy Evolution Explorer (GALEX) FUV map and the \textit{Spitzer} MIPS 24\,$\mu$m, 70\,$\mu$m, and 160\,$\mu$m maps of M81 in Figure \ref{fig:multiwav-map}, from which we estimated the SFRs. To obtain FUV luminosities, we sky-background subtract the FUV map and correct for foreground dust extinction using a constant $A_V=0.266$ (the same value used for our CMD fits) and $A_{\mathrm{FUV}}/A_{V} = 2.63$ \citep[][for $R_V=3.1$]{Fitzpatrick1999}. Adopting instead the empirical GALEX foreground calibration of \citet{Peek2013}, as in \citet{Leroy2019}, would raise all FUV luminosities uniformly by $\sim20\%$. We obtain the TIR luminosities by combining the 24\,$\mu$m, 70\,$\mu$m, and 160\,$\mu$m luminosities following \citet{D&H2002}.

\begin{figure}
    \centering
    \includegraphics[width=\linewidth]{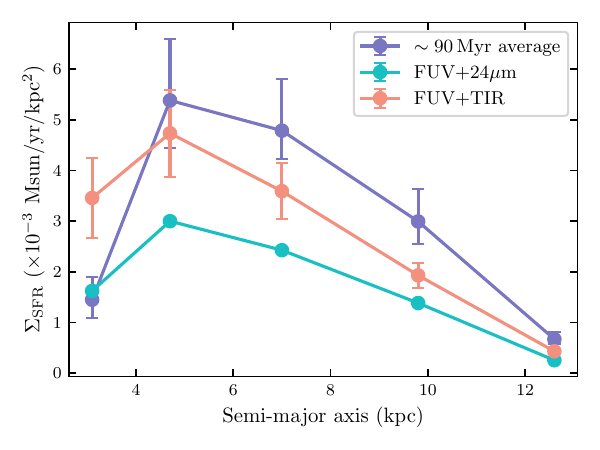}
    \caption{Comparison of CMD-based $\sim90\,$Myr average SFR density (blue) with the SFR density derived by combining FUV+$24\mu$m (cyan) and FUV+TIR (salmon) luminosities as a function of radius in M81. The integrated-light SFRs assume a \citet{Kroupa2003} IMF; for a constant SFH, converting them to the \citet{Kroupa2001} IMF would increase them by $\sim13\%$.}
    \label{fig:cmd-hybrid-comparison}
\end{figure}

The CMD-based SFR density in each of our M81 shells, averaged over $\sim90\,$Myr, is systematically higher than the FUV luminosity-based SFR densities (Figure \ref{fig:cmd-hybrid-comparison}). Outside Shell 1, the FUV+TIR SFR densities are lower than the CMD SFR densities by $\sim10-30\%$, while the FUV+24$\mu$m SFR densities are lower by $\sim40-60\%$. In the innermost shell, the FUV+TIR SFR density is higher than the CMD SFR density by a factor of $\sim2$, consistent with these radii being where dust heating by older stellar populations is most likely to contribute to the TIR.

A similar trend holds for the global SFRs listed in Table \ref{tab:SFRs}. We see that the uncorrected FUV luminosity of M81 misses $\sim40-60\%$ of the star formation inferred after dust obscuration corrections. This fraction agrees with the work of \citet{Perez-Gonzalez2006}, who found that about $\sim50\%$ of the star formation in M81's HII regions is obscured by dust and seen in the IR, whereas the other half can be seen through H$\alpha$ emission. Our CMD-based global SFR also agrees well with the TIR-only SFR of $0.89\,M_\odot\,\mathrm{yr}^{-1}$ obtained by \citet{Gordon2004}.

The lower FUV+24$\mu$m SFRs, by a factor of $\sim$2, are consistent with an emerging trend in which integrated-light tracers fall short of CMD-based SFRs in nearby spirals. In M31, 24$\mu$m-corrected FUV underestimates the 100\,Myr-averaged CMD SFR by a factor of 2.3--2.5 \citep{Lewis2017}, and \citet{Wainer2026} find a factor of $\sim$2.1 over the full disk, attributing it to M31's declining SFR, which violates the constant-SFR assumption built into these calibrations. The size of this offset, however, depends on calibration choices. Correcting for the different high-mass IMF slopes and adopting the \citet{Peek2013} foreground extinction would together raise the FUV term by $\sim35\%$. This would reduce the FUV+24$\mu$m deficit to a factor of $\sim1.7$ and close most of the remaining gap between the FUV+TIR and CMD SFRs outside Shell 1. The sense of the offset is therefore robust, but its amplitude is uncertain at the tens-of-percent level.

\begin{deluxetable}{lc}
\tablecaption{Global Star Formation Rate of M81\label{tab:SFRs}}
\tablehead{
\colhead{Tracer} & \colhead{SFR ($M_\odot\,\mathrm{yr}^{-1}$)}
}
\startdata
CMD     & $1.06^{+0.18}_{-0.20}$ \\
CMD (finite mass limits) & $0.80^{+0.13}_{-0.15}$ \\
$L_{\mathrm{FUV, obs}}$ & $0.249^{+0.001}_{-0.001}$ \\
$L_{\mathrm{FUV,obs}}+ 3.89^{0.15}_{0.15}\times L_{24\mu\mathrm{m}}$ & $0.41^{0.01}_{0.01}$ \\
$L_{\mathrm{FUV,obs}}+ 0.46^{0.12}_{0.12}\times L_{\mathrm{TIR}}$ & $0.6^{0.1}_{0.1}$ \\
\enddata
\tablecomments{The SFR from \texttt{MATCH} has to be corrected by -0.12\,dex to make it compatible with SFRs obtained by integrating the IMF with finite mass limits. The integrated-light SFRs assume a \citet{Kroupa2003} IMF; for a constant SFH, converting them to the \citet{Kroupa2001} IMF used for the CMD SFRs would increase them by $\sim13\%$. (Section~\ref{sec:sfr-indicators}).}
\end{deluxetable}

\section{Understanding spatial trends}\label{sec:physics}
The general increase in SFR surface densities towards lower galactocentric radii is the expected baseline for M81. This inward increase reflects the fact that the SFR density is ultimately set by the local gas supply and the local gravitational potential--- both of which increase toward the center in a normal exponential disk. This is thus a direct manifestation of the resolved Kennicutt-Schmidt relation; the molecular gas in the disk becomes more centrally concentrated and efficient at forming stars, driving up the SFR density \citep[][]{Kennicutt1998, Bigiel2008, Leroy2008}. The drop in SFR densities $\lesssim4\,$kpc is the deviation from this norm, which we now explore.

\subsection{A sign of AGN feedback?}
M81 is the nearest spiral galaxy with a compact radio source at its center, and with Centaurus A, is the nearest galaxy with an active galactic nucleus (AGN). The AGN in M81 is powered by a supermassive black hole (SMBH) of mass $4.78\times10^{7}\,M_\odot$ \citep{Nguyen2026} and is a low luminosity AGN (LLAGN) with properties intermediate to a radio-loud AGN and Sgr\,A$^*$ \citep{Markoff2008, Goold2026}. So, could M81's AGN be responsible for suppressing star formation in the central regions in the last $\sim100\,$Myr?

It has long been established that AGN feedback plays a crucial role in regulating star formation in galaxies \citep[e.g.,][]{Fabian2012}. Specifically, negative AGN feedback, or the suppression of star formation, can occur through two channels. Ejective feedback occurs when winds or jets from the AGN expel or displace the gas that would otherwise form stars. Preventive feedback occurs when these winds or jets primarily serve to keep the surrounding gas hot and turbulent, preventing it from cooling, condensing, and forming stars. Both feedback channels can operate in tandem in high accretion rate (radiatively efficient) and low accretion rate (radiatively inefficient) AGN through multiple overlapping mechanisms. However, in a radiatively inefficient LLAGN like M81, feedback is likely facilitated through jets launched from the central engine \citep{Alexander2025, Harrison&Ramos-Almeida2024}.  

M81's AGN is currently driving both a one-sided, sub-parsec scale radio jet \citep{Bietenholz2000} and an energetic hot wind \citep{Shi2021}. Further, M81's central region exhibits significant diffuse X-ray emission, distinct from resolved X-ray binaries, which is spatially concentrated in the bulge and well described by a soft thermal plasma component \citep{Markoff2008, Swartz2003, Li2010}. The diffuse hot gas may be partially powered by AGN feedback, as winds and jets from the central black hole can inject energy into the surrounding medium, heating and disturbing the X-ray–emitting plasma. These considerations motivate a picture where M81's AGN could be driving contemporary feedback through the preventative mode. However, due to a lack of temporal information regarding the AGN and the surrounding ISM, it is challenging to establish a direct causal link between M81's historical central quiescence and its SMBH.

\subsection{Secular Evolution}\label{sec:ILR}
M81's central quiescence could also reflect secular evolution tied to its spiral structure and classical bulge. In the classical density wave picture, spiral arms have a fixed pattern speed with well-defined corotation and resonance regions. Inside corotation, gas orbits faster than the pattern and shocks as it enters the arms, losing energy and drifting toward the center. In a galaxy with a classical bulge like M81, the arms become azimuthal at the ILR ($\sim4\,$kpc in the literature) and terminate there. As the pitch angle of the arm and its shock approaches $0\degr$, gas meets the shock front at increasingly oblique angles and its energy loss rate drops, stalling the inflow at the ILR and starving the region interior to it of molecular gas
\citep{Kormendy2004}.

This picture may not apply cleanly to M81, however, because its spiral arms are tidally induced \citep[e.g.,][]{Visser1980a, Visser1980b, Elmegreen1989,
Choi2015}. Simulations suggest that tidal arms behave in a manner intermediate between dynamic and density-wave-like structures: their pattern speed typically decreases with radius and lies below the material speed, winding the arms up over 2--3 disc rotations \citep[e.g.,][]{Pettitt2017, Oh2015}. A single, globally fixed ILR is therefore difficult to define, though a nominal ILR radius still follows from the local pattern speed. This falls near $\sim4\,$kpc given M81's centrally concentrated bulge. Because the pattern speed remains below the material speed, gas can still lose energy at the arms and drift inward rather than stalling sharply at this radius. Separately, \citet{Pettitt2017} note that the bulge can suppress central star formation efficiency as the randomized orbits of bulge stars can stabilize the inner gas disc against collapse. Reduced central star formation is thus plausible in either framework, albeit through different mechanisms.

Our observations are consistent with this picture but cannot isolate either mechanism. Molecular gas observations extend only to $\sim1.6\,$kpc \citep{Casasola2007}, so we constrain only the neutral gas: the present-day H{\sc i} map shows virtually no emission at $\lesssim4\,$kpc, while the $\sim25$--$100\,$Myr stellar map shows a ring-like structure near the nominal ILR (Figure \ref{fig:age-maps}), plausibly marking where inflowing gas accumulated and formed stars. The H{\sc i} deficit is equally consistent with gas having stalled and starved the interior or having flowed through, become molecular, and been stabilized against collapse. Either process, or both, could contribute to the low central star formation we recover over the last $\lesssim100\,$Myr.

\subsection{East-West Asymmetry}\label{sec:dust}
In Section \ref{sec:results-asymmetries} we found that the eastern half of M81 formed stars at a $\sim50\%$ higher rate than the western half over the last $\sim90\,$Myr, an asymmetry significant at the $>1\sigma$ level. Intriguingly, the integrated-light view inverts this result. The FUV luminosity of the western half is $\sim37\%$ higher than that of the eastern half---opposite to the resolved-star SFR asymmetry. This persists even after correcting the FUV with $24\,\mu$m emission, leaving a $\sim26\%$ spurious western excess. Taken at face value, an FUV-based SFR estimate would not only miss the eastern excess but flip its sign.

Crucially, such an inversion is not observed between the north and south halves. The FUV luminosity for the northern half is $\sim13\%$ higher than the southern half, respecting the sign of the CMD-inferred north-south asymmetry ($\sim18\%$). Thus, the east-west inversion appears to be the signature of a line-of-sight dust asymmetry between the near (western) and far (eastern) sides of M81's inclined disk---proportionately affecting the FUV luminosities of north and south halves while disproportionately affecting the east and west halves. In such a scenario, the young stars responsible for the FUV emission on the near side are, on average, seen through a smaller dust column than their far-side counterparts, so the near side is less attenuated in the FUV. This hypothesis is independently supported by the modestly lower differential extinction ($dA_V$) we recover in the western half. We do not attempt to model the detailed dust geometry responsible for this offset. In M81 the spatial relationship between the young stars and the dust downstream of the spiral shock is complex, with both distributed over a broad spiral compression zone \citep{Kaufman1989}. What matters here is the sense of the asymmetry, which both the $dA_V$ values and the FUV excess independently support.

This demonstrates a fundamental limitation of FUV-based SFR indicators in inclined galaxies. Even with an infrared dust correction, an orientation-dependent obscuration asymmetry can survive at the tens-of-percent level and, in this case, reverse the true sign of an internal SFR contrast. It underscores the value of resolved stellar photometry, which circumvents dust-induced flux asymmetries by directly counting the stars responsible for recent star formation.

\section{Comparisons with other $L^*$ galaxies}
M81 is only the third non-Milky Way $L^*$ galaxy after M31 and NGC6946 for which spatially-resolved SFHs have been derived from resolved stars. Here, we compare M81's radial and global SFH trends with those of M31 and NGC 6946.

Throughout the last $\sim180\,$Myr, M81's star formation has been dominated by star formation occurring between $\sim3.8-8.4\,$kpc. In contrast, most of M31's star formation in this time span has occurred at higher galactic radii in the $\sim10\,$kpc ring, with a modest amount occurring in an inner $\sim5\,$kpc ring \citep[][]{Lewis2015, Wainer2026}. In both M81 and M31, which host prominent central bulges, star formation over the last $\sim100\,$Myr is suppressed in the innermost $\sim4\,$kpc. In the circumnuclear starburst galaxy---NGC6946---on the other hand, there is no clear radial trend in star formation in the last $\sim25\,$Myr \citep{Tran2023}.

Looking at global numbers, the SFR in M31 is smaller than that in M81 by a factor of 2-6 in every age bin $\lesssim180\,$Myr. Unlike M81, which has a roughly constant SFH with slight enhancements $\lesssim15\,$Myr, M31's SFR declines by a factor of 4-5 over the last $\sim100\,$Myr, with the steepest decline occurring in the last $\sim40\,$Myr. In contrast, the SFR in NGC 6946 is up to $\sim10$ times larger than M81 in the last $\sim25\,$Myr. However, there is a global decline in SFRs by a factor of $\sim5$ in the last $\sim25\,$Myr. Thus, M81's SFH is markedly different from the declining SFHs of M31 and NGC 6946, highlighting the diverse evolutionary paths similar looking galaxies can take depending on their environments.

Large-scale asymmetries in star formation appear to be a common feature among nearby $L^*$ galaxies. In M81, we saw that the eastern half formed $\sim50\%$ more stars than the western half in the last $\sim90\,$Myr in the same area. In M31, the southern half formed $\sim20\%$ more stars than the northern half in the last $\sim50\,$Myr in the same area \citep{Lewis2015, Wainer2026}. Though the degree of asymmetry vanishes beyond $\sim100\,$Myr, it is challenging to fully quantify the true scale of the asymmetry due to limited spatial coverage of the SFH maps. In NGC 6946, the northern half of its disk formed $\sim1.5$ times more stars than the southern half and its eastern half formed $\sim1.2$ times more stars than its western half in the last $\sim16\,$Myr \citep{Tran2023}. From integrated light observations, lopsidedness has been established to be a generic feature of nearby spirals \citep{Zaritsky2013}. \citet{Z&R1997} find that $\sim30\%$ of their sample field spiral galaxies are lopsided whereas \citet{Bournaud2002} find that $\sim63\%$ of their sample are lopsided. The latter's numerical simulations also show that such asymmetries can be induced through tidal interactions, mergers, and asymmetric accretion of cosmic gas. The resolved star evidence thus adds a dust-independent line of support to the conclusion that lopsidedness is not an anomaly but a defining characteristic of $L^*$ spiral disks---one that likely reflects the ubiquity of the tidal interactions, minor mergers, and asymmetric gas accretion that drive it.

\section{Constraining M81's interaction history}\label{sec:merger-history}
\begin{figure}
    \centering
    \includegraphics[width=\linewidth]{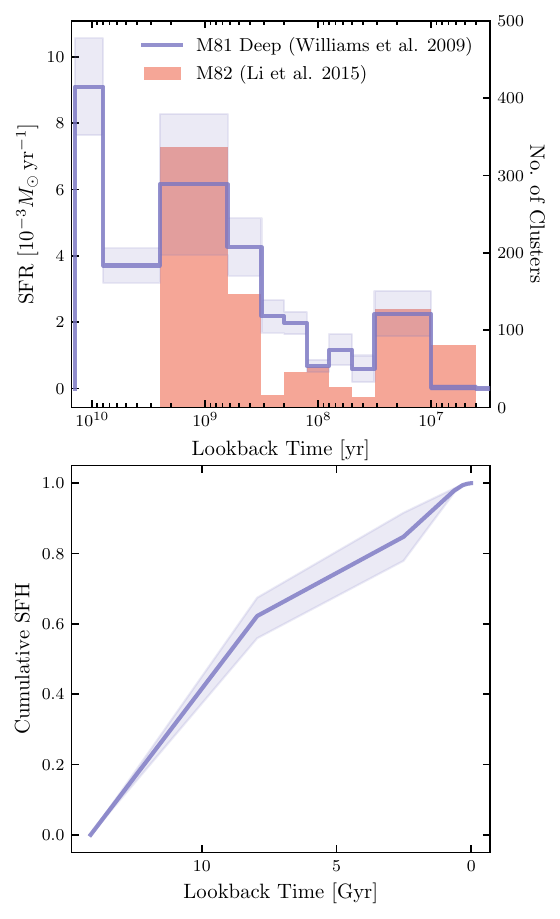}
    \caption{(Top) SFR as a function of lookback time in the M81-DEEP outer-disk field, located $\sim16\,$kpc to the north of M81 along the major axis \citep[blue; reproduced from ][]{Williams2009}. We compare this with the age distribution of star clusters in M82 using the same time bins \citep[salmon; reproduced from ][]{Li2015}. (Bottom) Cumulative SFH of the M81-DEEP field. Prominent stellar mass buildup events occur $\lesssim7.9\,$Gyr and $\sim1\,$Gyr.}
    \label{fig:m81-deep-sfh}
\end{figure}
The timing of M81's last close interaction with M82 is uncertain. One numerical N-body simulation of the M81 Group places the timing of last pericenter between M81 and M82 at $510\,$Myr \citep{Brouillet1991}, whereas another places it at $220\,$Myr \citep{Yun1999}. These simulations were primarily constrained by the H{\sc i} morphology of the group available at the time \citep[e.g.,][]{Yun1994} and do not incorporate the SFH evidence of either group members. On the other hand, some integrated light studies of M82, favor older pericenter times around $\sim1\,$Gyr \citep{deGrijs2001, Mayya2006}, whereas other star cluster studies \citep{Smith2007, Li2015} point to a wider range of times from $\sim150\,$Myr to $\sim1\,$Gyr. In this section, we bring the existing SFH evidence of M81 to bear upon this problem and use the IllustrisTNG collaboration's TNG50 cosmological simulation \citep{Nelson2019, Pillepich2019} to strengthen our constraints.

In M81, we have seen that star formation is nearly constant over the last $\sim180\,$Myr. We complement this recent SFH with the ancient SFH of M81's outer disk obtained by \citet{Williams2009}. The outer disk field (M81-DEEP) is $\sim16\,$kpc from M81 along the major axis to the north. It contains a branch of the western spiral arm and is outside our survey footprint. The ACS photometry for this field extends one magnitude fainter than the ancient red clump, and hence allows SFH determinations up to ancient times. M81-DEEP's SFH shows that most of the stellar mass was built up by a lookback time of $\sim7.9\,$Gyr ($z\sim1$; Figure \ref{fig:m81-deep-sfh}) and there was an uptick in star formation around $\sim1\,$Gyr. Following this, the SFR declined by a factor of $\sim4-6$ by $\sim100\,$Myr and increased slightly $\lesssim30\,$Myr. The cumulative SFH shows that the stellar mass build up events $\lesssim7.9\,$Gyr and $\sim1\,$Gyr were, by far, the dominant events in M81's past. From M31, we know that while long-term trends in SFH can vary as a function of radius, the dominant global features---particularly those corresponding to the main episodes of stellar mass buildup---are recoverable even from small-area SFHs \citep{Williams2015, Williams2017}. We therefore assume that the long-term SFH derived for the M81-DEEP field is qualitatively representative of M81's stellar disk as a whole.

The star formation in M82 also peaked $\sim1\,$Gyr ago (Figure \ref{fig:m81-deep-sfh}). We trace M82's SFH using the age distribution of star clusters from \citet{Li2015}'s catalog of 846 star clusters. These clusters were aged using multiple passband HST images spanning the entire optical and near IR wavelength range. The M82 star cluster age distribution shows a prominent peak $\sim1\,$Gyr ago, coinciding with the timing of the SFR enhancement in the M81 disk. Additionally, there are smaller peaks $\sim100\,$Myr ago and $\sim10\,$Myr ago. 

Close tidal encounters and pericentric passages can enhance star formation in both satellites and hosts via tidal torques, gas compression, and induced inflows, with the strongest effects typically occurring near first pericenter \citep[e.g.,][]{Scudder2012, Patton2013, diCintio2021}. Thus, the coincidence of enhanced star formation $\sim1\,$Gyr ago in both M81 and M82 is striking, and strongly suggests that an M81-M82 pericenter occurred around that time. While there is a slight increase in the distribution of M82 star cluster ages $\sim200-300\,$Myr, which coincides with the $220\,$Myr pericenter time predicted by \citet{Yun1999}, there is no corresponding increase in M81 star formation. Given the constant M81 SFH in the last $\sim150\,$Myr, it is unlikely that a pericenter occurred $220\,$Myr ago.

\begin{figure}
    \centering
    \includegraphics[width=\linewidth]{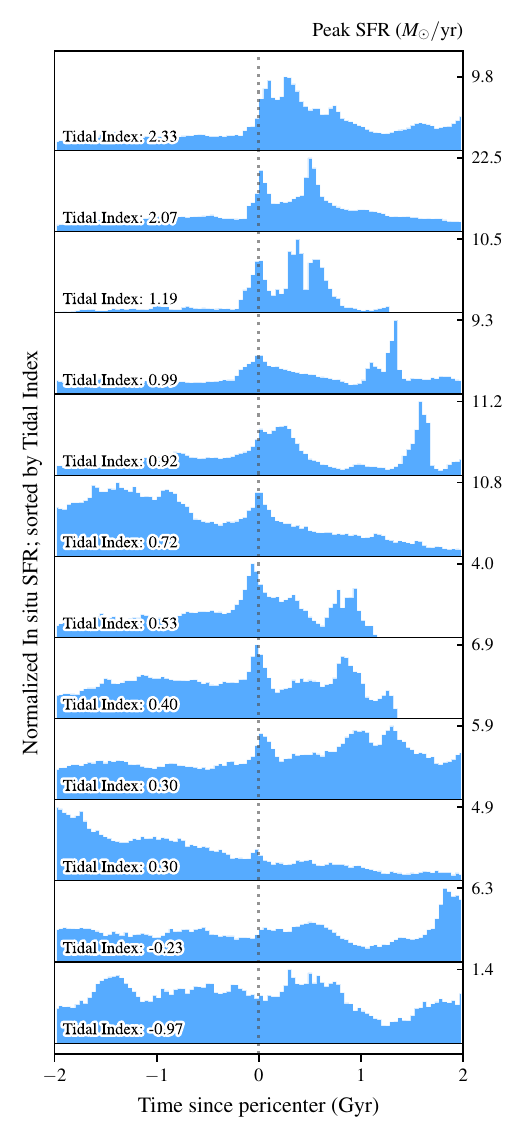}
    \caption{Normalized in-situ SFR vs. time of massive galaxies in TNG50 that undergo a pericenter passage by a comparably-massed galaxy (see text for details). Top to bottom, the plots are ordered by tidal index at first pericenter. Tidal index is the log ratio of tidal force exerted by the satellite ($\sim M_{\mathrm{sat}}/r^3$) to M82's current tidal force (assuming $M\sim3\times10^{10}\,M_\odot$ and $r\sim45\,$kpc). Peak SFRs in $M_\odot/\mathrm{yr}$ are also shown to the right of each panel to help readers judge the relative intensities of SFR fluctuations.}
    \label{fig:tng}
\end{figure}

To validate these constraints further, we look for M81-M82 analogs in the TNG50 simulation \citep{Nelson2019,Pillepich2019} to understand what the SFR of an M81-like galaxy should look like after a pericentric passage by an M82-like galaxy. Such interactions are not particularly common, so we allow a wide range of primary and secondary stellar masses to ensure a reasonable sample size. We choose all those systems with a primary galaxy mass $10^{10}M_{\odot} \le M_{*,\mathrm{primary}} \le 2\times 10^{11} M_{\odot}$, dark matter halo mass $<10^{13}M_{\odot}$, a merging secondary between  $3\times10^{9}M_{\odot} \le M_{*,\mathrm{secondary}} \le 5\times 10^{10} M_{\odot}$, a time of first pericenter passage between 0.5 and 4\,Gyr before the snapshot corresponding to the present day, and consider only star forming primaries with a mean SFR in the 4\,Gyr around the first pericenter time of $>0.3M_{\odot}\mathrm{yr^{-1}}$. The resulting SFRs of the twelve systems satisfying these criteria are shown in Fig.\ \ref{fig:tng}, offset vertically for clarity, and ordered by the strength of `tidal index' $\log_{10} (M_{*,\mathrm{secondary}}/3\times10^{10}M_{\odot}) (r_{\mathrm{pericenter}}/45\,{\mathrm{kpc}})^{-3}$, referenced to M82's approximate stellar mass and current projected distance to M81. 

The sample of TNG50 analogs show that any interaction with a tidal force even modestly greater than M82's current tidal force necessarily induces a burst in star formation in an M81-like galaxy lasting at least $\sim200-300\,$Myr (Figure \ref{fig:tng}). This comparison further motivates arguments that M82 experienced a pericenter $\sim1\,$Gyr ago since we see a substantial stellar mass buildup in the M81 disk around that time. Interestingly, most of the subsequent increases in SFR in the host occur due to following pericenter passages, or coalescence of the two merger partners with an accompanying starburst. This suggests that a $150-220\,$Myr M82 pericenter time \citep{Yun1999, Li2015} is incompatible with the constant M81 SFH observed over the last $\sim150\,$Myr. Instead, our results are more in line with the earlier predictions of \citet{deGrijs2001} and \citet{Mayya2006}.

M82 is currently moving with a line-of-sight velocity of $308\pm5\,\mathrm{km\,s^{-1}}$ with respect to M81 \citep{Speights2012, vanderTak2008}. From ram-pressure stripping considerations, \citet{Rao2026} propose an on-sky velocity of $\sim30-50\,\mathrm{km\,s^{-1}}$ for M82, making its relative velocity with M81 $\sim309-312\,\mathrm{km\,s^{-1}}$. Satellite galaxies orbiting Milky Way–mass hosts typically reach velocities of $\sim200–400\,\mathrm{km\,s^{-1}}$ near pericenter and $\sim50–150\,\mathrm{km\,s^{-1}}$ near apocenter, with exact values depending on orbital eccentricities \citep[e.g.,][]{Santisteven2023}. This suggests that M82 is likely near pericenter since the velocity near apocenter would be lower. These considerations imply that either M82 is currently moving away from a $\sim1\,$Gyr pericenter or will pericenter again in the future. If M82 is going out from a pericenter $\sim1\,$Gyr ago, it must be presently substantially farther than $\sim300\,$kpc from M81 as its velocity from pericenter to present would have been $\gtrsim300\,\mathrm{km\,s^{-1}}$. With current projected distances $\lesssim50\,$kpc, a substantial line-of-sight distance offset would be measurable. We thus conclude that M82 is likely approaching a pericenter with M81.

We have thus far ignored NGC 3077---a lower mass counterpart of M82 \citep[e.g.,][]{Walter2002}---from our consideration of M81's merger history. The same N-body simulations from \citet{Yun1999} predict that NGC3077 pericentered $280\,$Myr ago. However, \citet{Okamoto2023} constrained NGC 3077's pericenter time with M81 to also be $\sim1\,$Gyr ago. They estimated this timescale by comparing structural features of its stellar halo to simulations that investigated how such structures evolve during tidal interactions \citep{Penarrubia2009}. Thus, it is possible that NGC 3077 pericentered around the same time as M82 $\sim1\,$Gyr ago, producing coeval effects on the M81 disk.

\section{Future Work}
We have seen above that $\sim1\,$Gyr ago was an important time period in the evolutionary history of the M81 Group. However, the M81 SFH and the M82 star cluster ages rather loosely constrain the M81-M82 pericenter time between $\sim630\,$Myr to $\sim2.5\,$Gyr ago. Moreover, our recent M81 SFH only goes back up to $\sim180\,$Myr. Detailed SFHs of the M81 disk beyond $\sim180\,$Myr ago with improved time resolution along with resolved star SFH measurements of the M82 disk will better help constrain the pericenter time.

Upcoming results from the Cibola NIRCam survey of resolved stars in the M82 disk will elucidate M82's SFH up to ancient times (Smercina et al, \textit{in prep.}). This will be made possible primarily through JWST’s superior resolution, depth, and dust-penetrating capabilities, enabling clear detection of the ancient red clump feature. However, if the pericenter constraints are to be improved, the Cibola survey must be complemented by a deep M81 resolved star survey that improves time resolution of its SFH. If such a survey captures substantial portions of the M81 disk, it would also enable mapping the spatial distribution of stellar populations older than $\sim180\,$Myr, testing whether these older populations continue to trace spiral structure, and constraining radial SFH trends over longer timescales. Together, these observations would provide a critical empirical framework for interpreting the long-term evolution of tidally induced spiral density waves, radial migration, and disk restructuring, offering essential observational constraints to complement and refine theoretical models and simulations of galaxy evolution.

\section{Summary}
In this paper, we measured the spatially-resolved SFH of much of the M81 disk in the last $\sim180\,$Myr using resolved-star photometry. Our results can be summarized as follows:
\begin{itemize}
    \item We present for the first time a panoramic resolved star map of the M81 disk using photometry from the ANGST survey. Stars younger than $\sim25\,$Myr closely follow the H{\sc i} distribution in clear two-armed spiral features. These features also show up in maps of stars as old as $\sim150\,$Myr.
    \item M81's SFH is mostly constant over the last $\sim180\,$Myr (average SFR: $1.06^{+0.18}_{-0.20}\,M_\odot\,\mathrm{yr}^{-1}$ in the last 90\,Myr) with slight enhancements seen $\lesssim15\,$Myr. This is markedly different from the SFHs of the two other $L^*$ galaxies---M31 and NGC 6946---for which spatially-resolved SFHs have been measured; both show globally declining SFRs in the recent past.
    \item M81's SFR surface densities generally increase towards the center of the galaxy and drop off in the inner $\sim4\,$kpc bulge-dominated region. This is likely due to gas starving within the ILR and/or M81's AGN influencing star formation in the inner regions.
    \item Radial trends in our CMD-derived average SFRs are in general agreement with SFRs inferred from a combination of FUV and IR luminosities. Our global CMD-based SFRs are systematically higher than the FUV$+24\mu$m SFRs by a factor of $\sim2$ and are consistent with the FUV+TIR SFRs within uncertainties. The sense of the FUV$+24\mu$m offset is robust, but its amplitude depends at the tens-of-percent level on the adopted IMF and foreground extinction corrections.
    \item The eastern half of M81 formed stars at a $\sim50\%$ higher rate than the western half over the last $\sim90\,$Myr---a trend that FUV-based indicators not only miss, but reverse, inferring $37\%$ more star formation in the west due to inclination-driven dust systematics. Even a $24\,\mu$m correction leaves a $\sim26\%$ spurious excess, underscoring the need for resolved stellar photometry to obtain unbiased SFR measurements in inclined galaxies.
    \item The SFH evidence for M81 and M82 strongly suggests a close pericenter passage $\sim1\,$Gyr ago---substantially earlier than the $150-220\,$Myr predicted by some works. The latter timescale is further disfavored by M81's nearly constant SFH over the last $\sim180\,$Myr, which shows no signature of the interaction-driven star formation enhancement predicted in galaxy formation simulations like TNG50, for galaxies who experienced such a recent pericenter passage.
\end{itemize}

\begin{acknowledgments}
We thank the anonymous reviewer for their comments, which improved the manuscript. This research is based on observations made with the NASA/ESA Hubble Space Telescope obtained from the Space Telescope Science Institute, which is operated by the Association of Universities for Research in Astronomy, Inc., under NASA contract NAS 5–26555. These observations are associated with program(s) 10584 \& 10250. This work was supported in part by NASA under grant number 80NSSC24K0084 under solicitation NNH22ZDA001N-ROMAN: Nancy Grace Roman Space Telescope Research and Support Participation Opportunities.
\end{acknowledgments}

\begin{contribution}
VVR was responsible for the analysis, development of the scientific interpretations, and writing. EFB performed the TNG50 analysis. Both EFB and AS were responsible for project ideation and mentoring, and were crucial for discussing scientific interpretations. TW helped in conducting comparisons with SFR indicators and offered insights regarding M31. BW helped with utilizing archival ANGST photometry and ASTs. All other authors contributed towards project execution, review, and editing.

\end{contribution}

%
\facilities{HST(STIS)}

\software{numpy \citep{numpy}, matplotlib \citep{matplotlib}, astropy \citep{astropy}, DOLPHOT \citep{Dolphin2000PHOT}, MATCH \citep{Dolphin2002, Dolphin2012, Dolphin2013, Dolphin2016}}


\appendix

\section{Metallicities}\label{sec:metallicities}
\begin{figure}
    \centering
    \includegraphics[width=\linewidth]{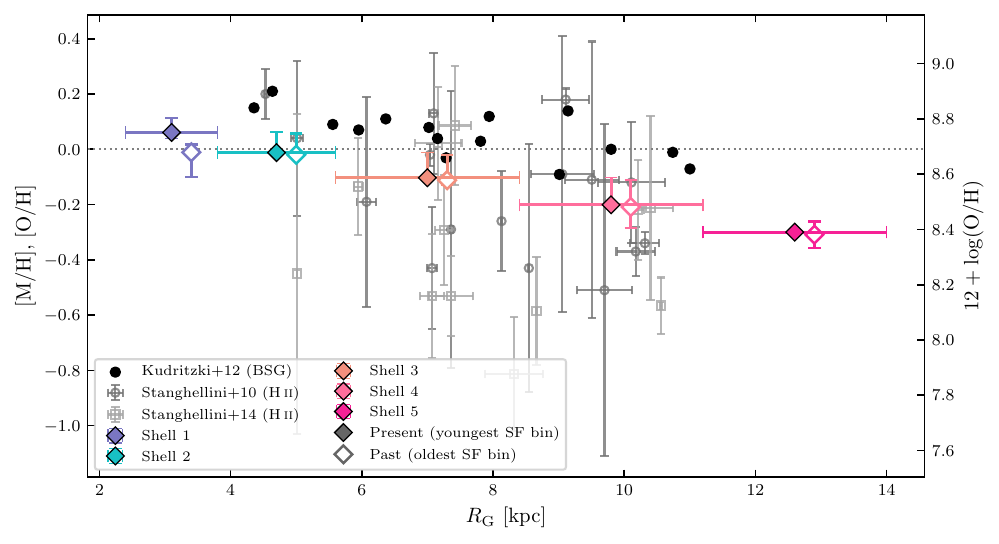}
    \caption{Radial metallicity profile of the M81 disk. Colored diamonds show the mass-weighted mean [M/H] from our CMD fits to each radial shell, plotted at the shell's central semi-major axis. Horizontal bars mark the radial extent of each shell, and shell colors match those in Figure~\ref{fig:sfh}. Filled diamonds give the present-day metallicity, taken from the youngest age bin with non-zero star formation ($\log t = 6.60$--$6.75$; $\sim$5\,Myr). Open diamonds give the past metallicity, taken from the oldest such bin ($\log t = 8.10$--$8.25$; $\sim$150\,Myr) and offset by $+0.3$\,kpc for clarity. Where the youngest or oldest bin has zero best-fit star formation, the adjacent bin is used instead. Vertical error bars combine random and systematic uncertainties. For comparison, black circles show blue supergiant (BSG) metallicities from \citet{Kudritzki2012}, and gray open circles and squares show direct-method H\,\textsc{ii}-region oxygen abundances from \citet{Stanghellini2010} and \citet{Stanghellini2014}, respectively. Oxygen abundances are placed on the [M/H] scale using a solar value of $12+\log(\mathrm{O/H})_\odot = 8.69$ \citep{Asplund2009}; the right-hand axis shows the native $12+\log(\mathrm{O/H})$ scale. All galactocentric radii are rescaled to our adopted distance modulus of $(m-M)_0 = 27.9$ (3.80\,Mpc). The dotted line marks solar metallicity.}
    \label{fig:mh-gradient}
\end{figure}

We show the past and present mass-weighted mean [M/H] from our CMD fits to each radial shell in Figure~\ref{fig:mh-gradient} and find negligible metal enrichment over the $\sim180\,$Myr time span. A noticeable metallicity increase with time is seen in our innermost shell, but we caution the reader from interpreting this outlier due to degeneracies with dust reddening. Rather remarkably, our CMD-derived metallicities are in broad agreement with independently derived blue supergiant (BSG) metallicities \citep[][]{Kudritzki2012} and H\,\textsc{ii} region nebular metallicites \citep[][]{Stanghellini2010, Stanghellini2014}. Our metallicity gradient of $-0.037 \pm 0.003\:\mathrm{dex/kpc}$ closely matches \citet{Kudritzki2012}'s $-0.033\pm 0.009\:\mathrm{dex/kpc}$. However, our [M/H] values are systematically lower than the BSG values by $0.17\:$dex. Both this work and \citet{Kudritzki2012} probe the same young population, but only our metallicities are degenerate with dust: in CMD fitting, reddening and metallicity both shift the colors of young stars, whereas \citet{Kudritzki2012} measure metallicity from continuum-normalized spectra and derive reddening separately for each star. We therefore treat their values as the more reliable reference.

\bibliography{main}{}
\bibliographystyle{aasjournalv7}



\end{document}